\documentclass[%
 reprint,
superscriptaddress,
 amsmath,amssymb,
 aps,prc,
]{revtex4-1}
\usepackage{multirow}
\usepackage{graphicx}% Include figure files
\usepackage{dcolumn}% Align table columns on decimal point
\usepackage{bm}% bold math
\usepackage{upgreek}
\usepackage[hidelinks]{hyperref}% add hypertext capabilities
\usepackage{CJKutf8}
\begin{document}
\preprint{APS/PLACEHOLDER}
\title{Identifying the spin trapped character of the  $^{32}$Si isomeric state}

\author{J.~Williams}
 \email{ewilliams@triumf.ca}
\affiliation{TRIUMF, 4004 Wesbrook Mall, Vancouver, British Columbia, Canada V6T 2A3}

\author{G.~Hackman}
\affiliation{TRIUMF, 4004 Wesbrook Mall, Vancouver, British Columbia, Canada V6T 2A3}

\author{K.~Starosta}
\affiliation{Department of Chemistry, Simon Fraser University, 8888 University Drive, Burnaby, British Columbia, Canada V5A 1S6}

\author{R.~S.~Lubna}
\affiliation{Facility for Rare Isotope Beams, Michigan State University, 640 South Shaw Lane, East Lansing, MI 48824}

\author{Priyanka~Choudhary}
\author{P.~C.~Srivastava}
\affiliation{Department of Physics, Indian Institute of Technology Roorkee, Roorkee, India 247667}

\author{C.~Andreoiu}
\author{D.~Annen}
\affiliation{Department of Chemistry, Simon Fraser University, 8888 University Drive, Burnaby, British Columbia, Canada V5A 1S6}
\author{H.~Asch}
\affiliation{Department of Physics, Simon Fraser University, 8888 University Drive, Burnaby, British Columbia, Canada V5A 1S6}

\author{M.~D.~H.~K.~G.~Badanage}
\affiliation{Department of Chemistry, Simon Fraser University, 8888 University Drive, Burnaby, British Columbia, Canada V5A 1S6}

\author{G.~C.~Ball}
\affiliation{TRIUMF, 4004 Wesbrook Mall, Vancouver, British Columbia, Canada V6T 2A3}

\author{M.~Beuschlein}
\affiliation{Technische Universit{\"a}t Darmstadt, Department of Physics, Institute for Nuclear Physics, Schlossgartenstr. 9, 64289 Darmstadt, Germany}

\author{H.~Bidaman}
\author{V.~Bildstein}
\author{R.~Coleman}
\affiliation{University of Guelph, 50 Stone Rd E, Guelph, Ontario, Canada N1G 2W1}

\author{A.~B.~Garnsworthy}
\affiliation{TRIUMF, 4004 Wesbrook Mall, Vancouver, British Columbia, Canada V6T 2A3}

\author{B.~Greaves}
\affiliation{University of Guelph, 50 Stone Rd E, Guelph, Ontario, Canada N1G 2W1}

\author{G.~Leckenby}
\affiliation{TRIUMF, 4004 Wesbrook Mall, Vancouver, British Columbia, Canada V6T 2A3}
\affiliation{Department of Physics and Astronomy, University of British Columbia, Vancouver, British Columbia, Canada V6T 1Z1}

\author{V.~Karayonchev}
\altaffiliation[Present address: ]{Argonne National Laboratory, 9700 S. Cass Avenue, Lemont, IL 60439}
\affiliation{TRIUMF, 4004 Wesbrook Mall, Vancouver, British Columbia, Canada V6T 2A3}

\author{M.~S.~Martin}
\affiliation{Department of Physics, Simon Fraser University, 8888 University Drive, Burnaby, British Columbia, Canada V5A 1S6}

\author{C.~Natzke}
\affiliation{TRIUMF, 4004 Wesbrook Mall, Vancouver, British Columbia, Canada V6T 2A3}

\author{C.~M.~Petrache}
\affiliation{Universit\'{e} Paris-Saclay, CNRS/IN2P3, IJCLab, 91405 Orsay, France}

\author{A.~Radich}
\affiliation{University of Guelph, 50 Stone Rd E, Guelph, Ontario, Canada N1G 2W1}

\author{E.~Raleigh-Smith}
\author{D. Rhodes}
\affiliation{TRIUMF, 4004 Wesbrook Mall, Vancouver, British Columbia, Canada V6T 2A3}

\author{R.~Russell}
\affiliation{University of Surrey, Guildford, United Kingdom GU2 7XH}

\author{M.~Satrazani}
\affiliation{University of Liverpool, Liverpool, United Kingdom L69 3BX}

\author{P.~Spagnoletti}
\affiliation{Department of Chemistry, Simon Fraser University, 8888 University Drive, Burnaby, British Columbia, Canada V5A 1S6}

\author{C.~E.~Svensson}
\affiliation{University of Guelph, 50 Stone Rd E, Guelph, Ontario, Canada N1G 2W1}

\author{D.~Tam}
\affiliation{Department of Physics, Simon Fraser University, 8888 University Drive, Burnaby, British Columbia, Canada V5A 1S6}
\author{F.~Wu (\begin{CJK*}{UTF8}{gbsn}吴桐安\end{CJK*})}
\affiliation{Department of Chemistry, Simon Fraser University, 8888 University Drive, Burnaby, British Columbia, Canada V5A 1S6}

\author{D.~Yates}
\affiliation{TRIUMF, 4004 Wesbrook Mall, Vancouver, British Columbia, Canada V6T 2A3}
\affiliation{Department of Physics and Astronomy, University of British Columbia, Vancouver, British Columbia, Canada V6T 1Z1}

\author{Z.~Yu}
\affiliation{Department of Chemistry, Simon Fraser University, 8888 University Drive, Burnaby, British Columbia, Canada V5A 1S6}

\date{\today}% It is always \today, today,
             %  but any date may be explicitly specified
\begin{abstract}
%Make sure to double check addresses of authors.
The properties of a nanosecond isomer in $^{32}$Si, disputed in previous studies, depend on the evolution of proton and neutron shell gaps near the `island of inversion'.  We have placed the isomer at 5505.2(2)~keV with $J^{\pi} = 5^-$, decaying primarily via an \emph{E3} transition to the $2^+_1$ state.  The \emph{E3} strength of 0.0841(10) W.u.~is unusually small and suggests that this isomer is dominated by the $(\nu d_{3/2})^{-1} \otimes (\nu f_{7/2})^{1}$ configuration, which is sensitive to the $N=20$ shell gap.  A newly observed $4^+_1$ state is placed at 5881.4(13)~keV; its energy is enhanced by the $Z=14$ subshell closure.  This indicates that the isomer is located in a `yrast trap', a feature rarely seen at low mass numbers.

%\begin{description}
%\item[Usage]
%Secondary publications and information retrieval purposes.
%\item[PACS numbers]
%May be entered using the \verb+\pacs{#1}+ command.
%\item[Structure]
%You may use the \texttt{description} environment to structure your abstract;
%use the optional argument of the \verb+\item+ command to give the category of each item. 
%\end{description}
\end{abstract}
%\pacs{Valid PACS appear here}% PACS, the Physics and Astronomy
                             % Classification Scheme.
%\keywords{Suggested keywords}%Use showkeys class option if keyword
                              %display desired
\maketitle
%\tableofcontents

\par Neutron-rich nuclides around the $N=20$ `island of inversion' contain rich information on nuclear structure.  For Na and Mg isotopes, residual nucleon-nucleon interactions lead to the dissapearance of the $N=20$ shell gap and substantial occupation of neutron $fp$ shell orbitals in the ground state configuration \cite{otsuka_tensorforce,ioi_theory,forces_utsuno}.  For nearby nuclides in the neutron-rich $sd$ shell, the evolution of the $N=20$ shell closure is indicated by intermediate energy negative parity states which arise mainly due to single neutron excitation to the higher-lying $fp$ orbitals.  Furthermore, the energies of normal (positive parity) configurations are affected by subshell closures within the $sd$ shell.  These differing excitation modes affect the energy spacing and ordering of normal and intruder configurations, influencing the decay scheme of the nucleus and potentially giving rise to nuclear isomerism.  Various studies have identified isomers near the `island of inversion' and have highlighted their importance for understanding the evolving structure in this region \cite{gray2023microsecond, 34alto34si_isomer, 32mal}.

\par For isomers in $sd$ shell nuclei, various systematic trends are apparent from existing data \cite{garg_atlas_2023}.  Spin isomers (where decays are hindered by large $\Delta J$ values, typically $\Delta J \geq 3$) are rare in this region and all occur in odd-$Z$ odd-$N$ nuclei near stability, except for two disputed cases in $^{26}$F and $^{32}$Si.  The case of $^{32}$Si is particularly interesting due to its proximity to the `island of inversion', as well as the fact that it is an even-$Z$ even-$N$ nucleus, and no spin isomers have been identified in even-even nuclei below $^{54}$Fe \cite{garg_atlas_2023}.  The scarcity of these isomers in low mass nuclei is due to the absence of intruder orbitals near the Fermi energy which would facilitate high spin states at low excitation energies.  For higher mass nuclei near shell closures, spin isomers are common due to the population of high-$j$ intruder orbitals such as $0g_{9/2}$ or $0h_{11/2}$ \cite{garg_atlas_2023}.  In principle, neutron cross-shell excitation to the $0f_{7/2}$ orbital could produce similar isomers in even-even $sd$ shell nuclei near $N=20$.  These isomers could arise either due to small energy gaps between normal and intruder states of similar spin which hinder transitions between those states, or due to the inversion of the yrast sequence such that a high spin state becomes lower in energy than lower spin states, resulting in a `yrast trap' that restricts the decay of the higher spin state.

\par These considerations have brought our attention to $^{32}$Si, the lowest mass even-even nucleus in which a spin isomer candidate has been reported.  Initially, a $5^- \rightarrow 4^+ \rightarrow 2^+_1$ cascade was proposed with a small $5^- \rightarrow 4^+$ decay energy leading to isomerism of the $5^-$ state with $\tau_{mean}\sim40$ ns \cite{32si_fornal}.  However, a subsequent report did not observe the $5^- \rightarrow 4^+$ transition and instead proposed a direct $5^- \rightarrow 2^+_1$ \emph{E3} decay \cite{32si_asai}.  The $5^- \rightarrow 4^+$ \emph{E1} strength reported in the former case would be similar to other transitions in this mass region, however if the latter case were true, it would imply some unusual properties for this isomer.  Notably, its \emph{E3} decay strength would be nearly an order of magnitude lower than any known \emph{E3} transition in the region, and its primary $\gamma$-decay energy would be the second largest of any known isomeric state, surpassed only by the decay of a core-excited \emph{E4} isomer in $^{98}$Cd \cite{98Cd_e4isomer}.  The disputed $^{32}$Si decay scheme has also cast into doubt the energy of its $4^+_1$ state, an important observable in even-even nuclei for testing {\it ab-initio} models.  A likely explanation for the isomerism of the $5^-$ state is that it is spin trapped due to being lower in energy than the $4^+_1$ state, however this could not be determined from the previous studies.  We report new experimental data confirming the inverted ordering of the $5^-_1$ and $4^+_1$ states in $^{32}$Si, making it the lowest mass even-even nucleus known to contain a spin trapped isomer.

\par Our experiment was performed at the ISAC-II facility of TRIUMF, where excited states of interest in $^{32}$Si were populated using a $^{12}$C($^{22}$Ne,2p) fusion-evaporation reaction with a $^{22}$Ne beam energy of 2.56A MeV.  The cross section of the $^{32}$Si channel was approximately 0.6 mb, corresponding to $\sim$1.5\% of the total data collected.  Gamma rays were detected using the TIGRESS array \cite{tigress} instrumented with 14 segmented HPGe clovers - 4 each at 45$^{\circ}$ and 135$^{\circ}$, and 6 at 90$^{\circ}$ with respect to the beam axis.  Charged particles were detected using a 128-channel spherical CsI(Tl) array \cite{csiball}.  Charged particle identification was performed using offline pulse shape analysis \cite{tip} to separate the 2-proton exit channel populating $^{32}$Si.  A self-supporting 500 $\upmu$g/cm$^2$ $^{nat.}$C target foil produced by Micromatter \cite{micromatter} was used with a 23.6 mg/cm$^2$ Pb catcher foil mounted approximately 1 mm downstream.  The purpose of the catcher foil was to stop $^{32}$Si recoils before the decay of the isomer, allowing separation of prompt and isomeric transitions based on the Doppler shift.  For lifetime measurements using the Doppler-Shift Attenuation Method (DSAM), an alternate target consisting of a 394 $\upmu$g/cm$^2$ layer of $^{\text{nat.}}$C on a 24 mg/cm$^2$ Pb backing was used \cite{targets}.

\par Spins of states populated in $^{32}$Si were determined from gamma-ray directional correlation ratio ($R_{DCO}$) values \cite{dco}, which were measured using two angular bins assigned to the TIGRESS clovers at 90$^{\circ}$, and the TIGRESS clovers at 45$^{\circ}$ or 135$^{\circ}$, respectively.  For these angles, values of $R_{DCO} \approx 0.5$ are obtained for stretched dipole transitions and $R_{DCO} \approx 1.0$ for stretched quadrupole or octupole transitions, when gating on a coincident stretched quadrupole transition.  Where sufficient statistics were available, the electric or magnetic character of transitions was determined using the polarization direction correlation method~\cite{STAROSTA199916}.  The measured polarization asymmetry $\Delta_{asym}$ was corrected for the intrinsic asymmetry in the response of TIGRESS measured using $^{56}$Co source data.  The validity of the above techniques was confirmed using transitions of known multipolarity in the $^{26}$Mg side channel.  Lifetimes of non-isomeric states were determined from a comparison of the DSAM target data to GEANT4-based simulations as described in Ref.~\cite{22ne} and previously implemented in Refs.~\cite{williams28mg,williams25na}.  Feeding corrections and estimations of systematic uncertainty due to the electronic stopping powers were performed using the methods of Ref.~\cite{williams25na}.

\par
\begin{figure}
\begin{center}
\includegraphics[width=1\columnwidth]{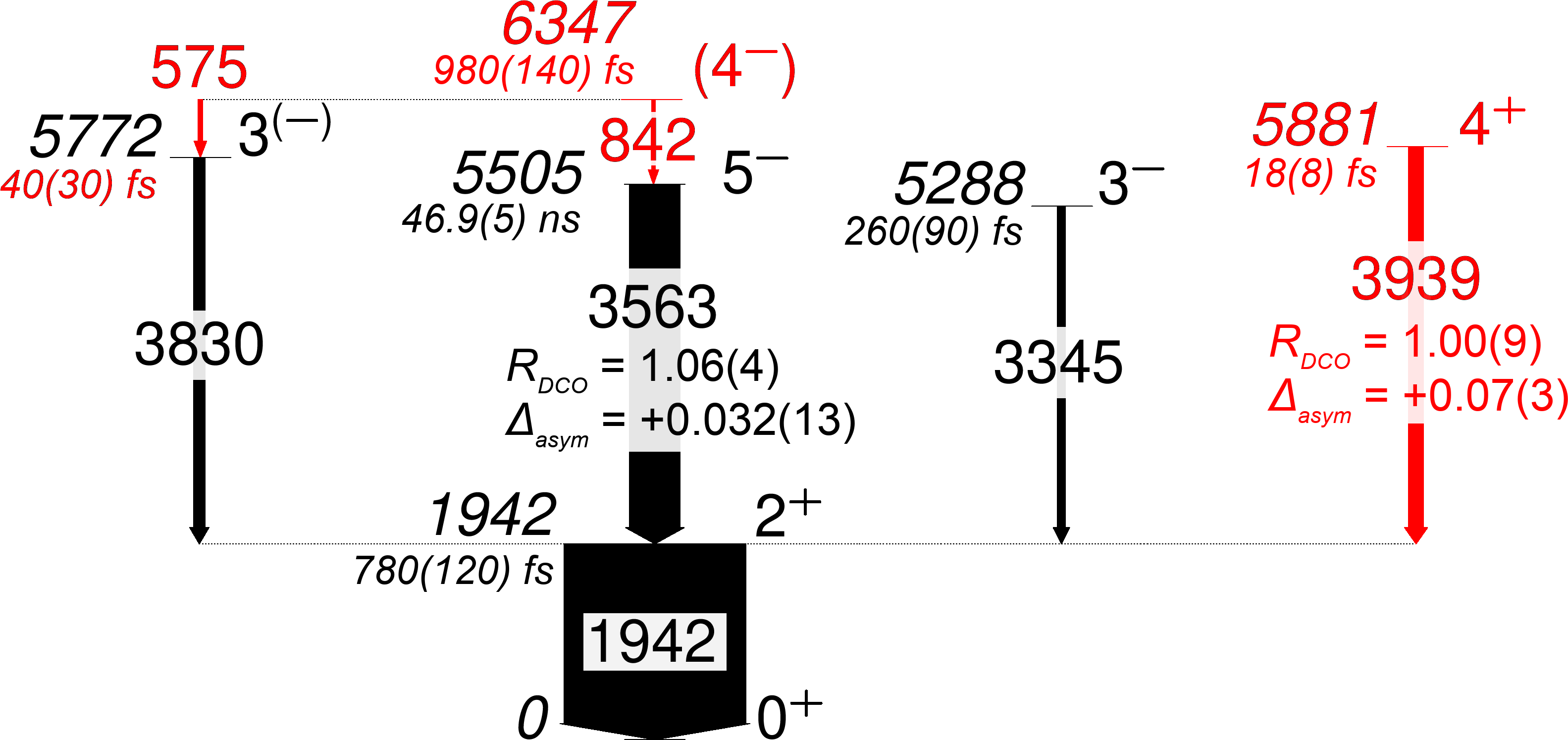}
\end{center}
\caption{Partial decay scheme of $^{32}$Si, with $R_{DCO}$ (gated on the $2^+_1 \rightarrow 0^+_1$ transition) and $\Delta_{asym}$ values listed for select transitions.  Line widths indicate relative intensities of each transition.  Newly observed levels and transitions and newly measured level lifetimes are in red.}
\label{fig:decayscheme}
\end{figure}

 A partial decay scheme for $^{32}$Si is shown in Figure \ref{fig:decayscheme}, with $R_{DCO}$ and $\Delta_{asym}$ values listed for select transitions, as well as the mean lifetime measured for each level.  Only levels and transitions necessary for identification and characterization of the isomeric state are shown. Additional states observed in this experiment will be presented in a future article.
%
%\begin{table}
%\caption{Partial list of $^{32}$Si levels observed in this work, and their corresponding $\gamma$ rays and lifetimes.  Items in {\bf bold} are newly observed or measured, items in {\it italic} are from Ref.~\cite{nds32}.}
%\centering
%\begin{ruledtabular}
%\begin{tabular}{llllr}
%\rule{0pt}{2.5ex}$E_{level}$ (keV) & $E_{\gamma}$ (keV) & $I_{\gamma, rel}$ & $\tau_{mean}$ (fs) & $J^{\pi}$ \\ \hline
%\rule{0pt}{2.5ex}1942.19(9) & 1942.13(9) & 100.0(5) & 780(120) & $\mathit{2^{+}}$\\
%5287.8(8)        & 3344.6(11) & 4.5(5) & 260(90) & $\mathit{3^{-}}$ \\
%5505.2(2)        & 3562.84(14) & 27.7(5) & $46.9(5) \times 10^6$ & $\mathbf{5^-}$ \\
%5771.6(13)       & 3829.9(12)       & 6.0(2) & {\bf 40(30)} & $\mathbf{3(^-)}$ \\
%{\bf 5881.4(13)} & {\bf 3938.9(13)} & 8.1(2) & {\bf 18(8)} & $\mathbf{4^+}$ \\
%{\bf 6347.3(4)} & {\bf 574.9(3)} & {\bf 2.3(3)} & {\bf 980(140)} & $\mathbf{(4^-)}$ \\
%                & {\bf 842.1(3)} & {\bf 1.7(2)} & &
%\end{tabular}
%\end{ruledtabular}
%\label{tab:32si_levels}
%\end{table}

%\begin{table}
%\caption{List of $a_2$, $a_4$, $R_{DCO}$ (gated on the $2^+_1 \rightarrow 0^+_1$ transition), and $\Delta_{asym}$ values measured for transitions in $^{32}$Si.  Transition multipolarities which could not be uniquely identified are assigned as dipole (D), quadrupole (Q), or both.}
%\centering
%\begin{ruledtabular}
%\begin{tabular}{lllllr}
%\rule{0pt}{2.5ex}$E_{\gamma}$ & \multirow{2}{*}{$a_2$} & \multirow{2}{*}{$a_4$} & \multirow{2}{*}{$R_{DCO}$} & \multirow{2}{*}{$\Delta_{asym}$} & Assign- \\ 
%(keV) & & & & & ment \\ \hline
%\rule{0pt}{2.5ex}3344.6(11) & -0.25(14) & 0.0(2) & 0.4(2) & $+0.05(8)$ & D \\
%3562.84(14) & 0.47(3) & -0.13(4) & 1.06(4) & $+0.032(13)$ & E3$^a$ \\
%3829.9(12) & -0.37(12) & -0.34(14) & 0.48(14) & $+0.04(7)$ & D \\
%3938.9(13) & 0.11(6) & -0.15(7) & 1.00(9) & $+0.07(3)$ & E2 \\
%574.9(3) & -0.3(3) & 0.1(3) & 0.9(2) & - & (D+)Q \\
%\rule{0pt}{2.5ex}842.1(3) & -0.5(7) & -0.7(8) &  0.6(2) & - & D(+Q) \\
%\end{tabular}
%\end{ruledtabular}
%{\footnotesize \rule{0pt}{2.5ex}$^a$From decay scheme and lifetime, see text.}
%\label{tab:32si_rdco_table}
%\end{table}

\par
\begin{figure}
\begin{center}
\includegraphics[width=1.0\columnwidth]{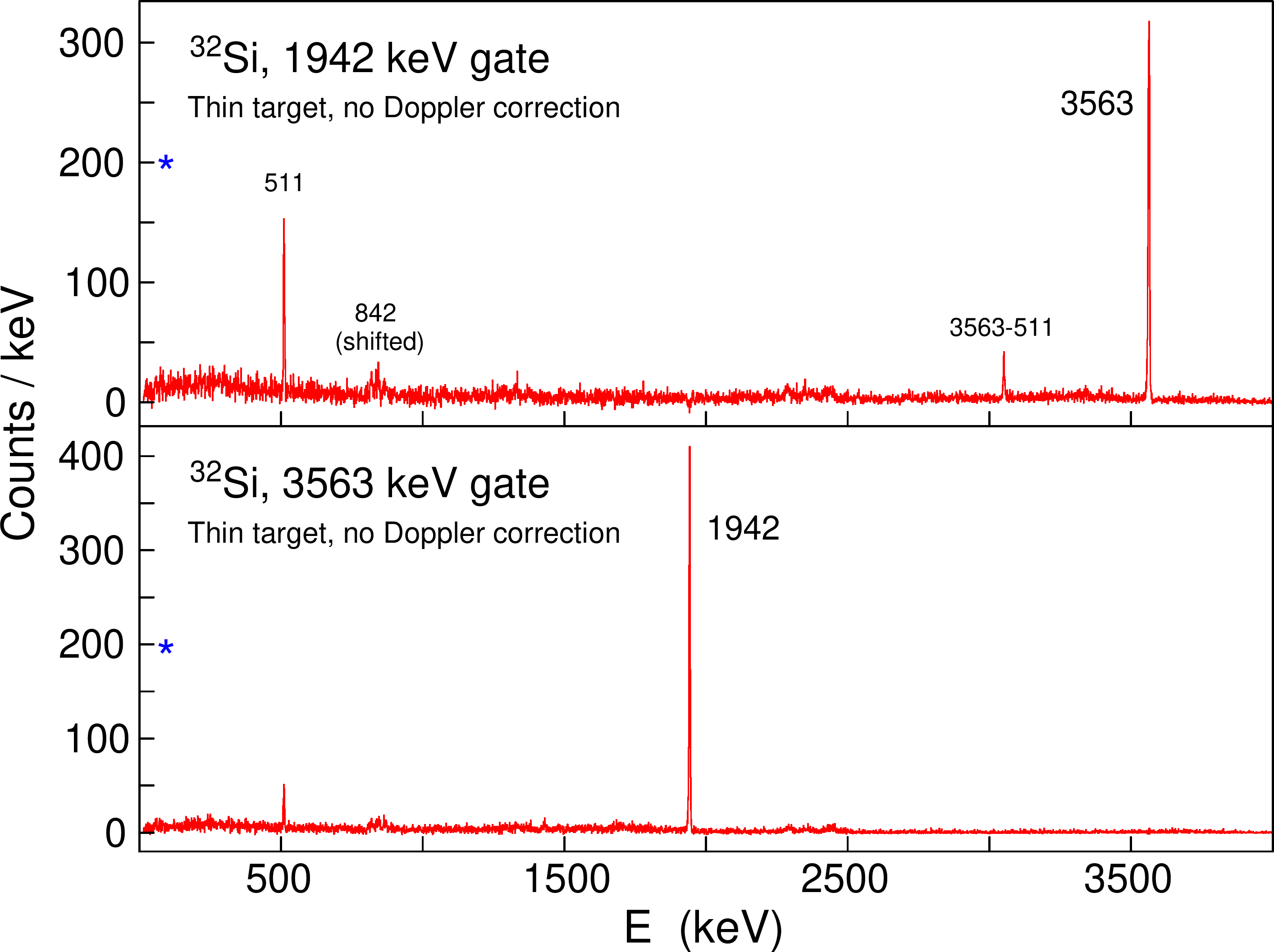}
\end{center}
\caption{Background subtracted gamma-ray spectra gated on stopped lines in $^{32}$Si.  The expected position and height of a 79~keV member of the isomeric cascade is labelled with the `*' character.}
\label{fig:EGamma_spectra}
\end{figure}

 The $^{32}$Si nanosecond isomer has previously been proposed at either 5581~keV \cite{32si_fornal} or 5504~keV \cite{32si_asai}.  In the former case, a 79(1)~keV transition was reported between the proposed isomeric state and the 5504~keV level.  In the present work we observe a delayed cascade depopulating a level at 5505.2(2)~keV, but no coincident 79~keV transition, see Figure \ref{fig:EGamma_spectra}.  The intensity upper limit of a hypothetical stopped line at this energy was determined to be 1.2\% relative to the fully stopped 3562.84(14)~keV line, based on the Compton background at 79~keV and the relative gamma-ray detection efficiency of TIGRESS at the energies of interest.
Prior to background subtraction, a stopped line at 78~keV was observed corresponding to time-random background from $^{32}$P, which is very strongly populated in the $^{12}$C($^{22}$Ne,pn) side channel.

\par Since the 79~keV transition of Ref.~\cite{32si_fornal} was not observed, we investigated the possible isomerism of the 5505.2(2)~keV level based on its spin-parity and the observed decay scheme.  The $R_{DCO}$ and $\Delta_{asym}$ values of Figure \ref{fig:decayscheme} strongly favor an \emph{E2} or \emph{E3} assignment for the 3562.84(14)~keV transition, consistent with the $J^{\pi} = (5^-,4^+)$ assignment for the parent level from previous $^{30}$Si$(t,p)$ data \cite{32si_fortune}.  Given the large transition energy, only the $5^-$ case would produce an isomeric state.
The excitation energy of the isomer was constrained based on the decays of a newly observed level at 6347.3(4)~keV, depopulated via a 842.1(3)~keV transition feeding the isomeric cascade, as well as a 574.9(3)~keV transition to a side band outside of the isomeric cascade.  Neither of the 574.9(3) or 842.1(3)~keV lines is a member of the isomeric cascade, as they are not stopped in the thin target data.  Comparing the total energies of each decay path from the 6347.3(4)~keV level to the ground state, the energy of the isomeric state is constrained to 5505.3(14)~keV, which agrees very well with the observed level at 5505.2(2)~keV.

\par
\begin{figure}
\begin{center}
\includegraphics[width=1.0\columnwidth]{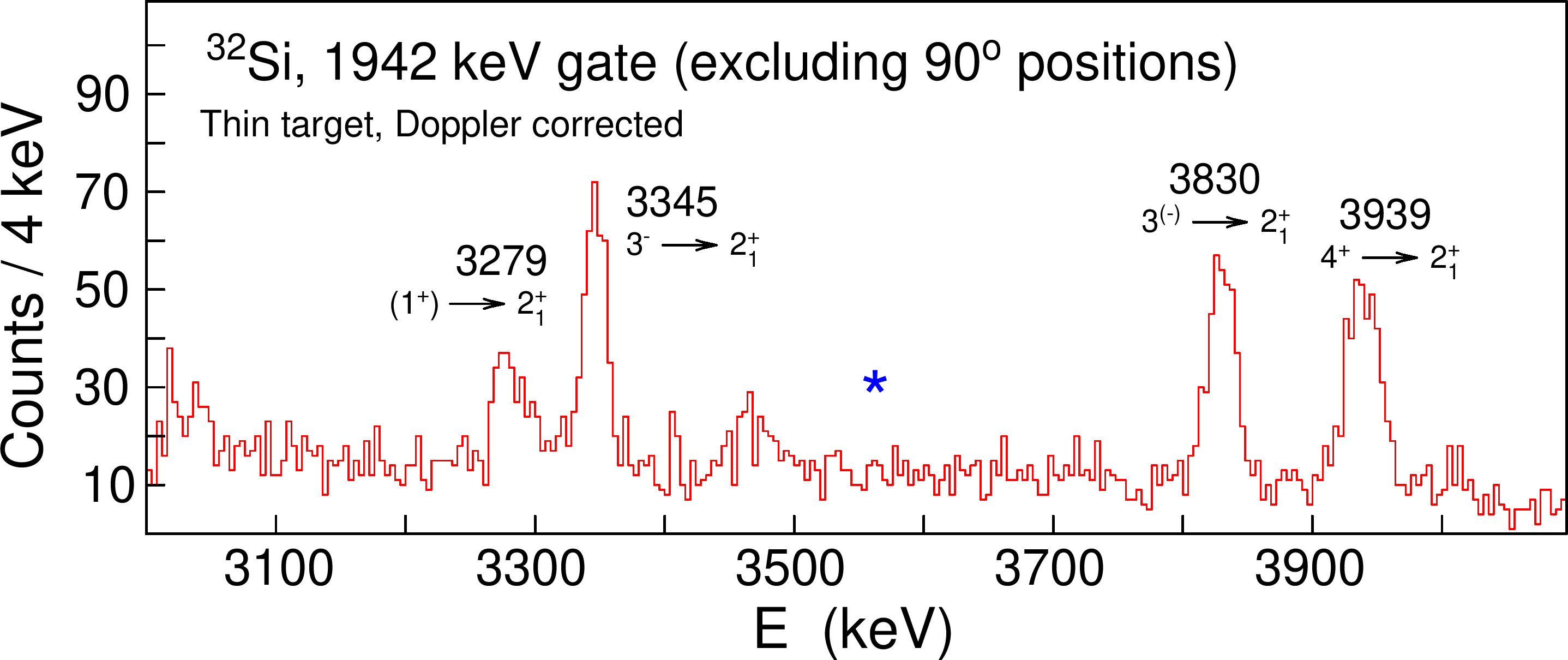}
\end{center}
\caption{Doppler corrected gamma-ray spectrum gated on the prompt component of the 1942.13(9)~keV transition in $^{32}$Si.  The expected position of a prompt component of the 3562.84(14)~keV transition is labelled with the `*' character.}
\label{fig:prompt3562}
\end{figure}

 Despite the observed prompt feeding of the 5505.2(2)~keV level, there was no observed prompt component of the 3562.84(14)~keV transition depopulating this level.  An upper limit on the intensity of an unobserved prompt component was estimated by gating on the Doppler shifted component of the $2^+_1 \rightarrow 0^+_1$ transition (excluding detectors at $90^{\circ}$ where the shifted and stopped energies overlap).  The resulting spectrum is shown in Figure \ref{fig:prompt3562} and contains no indication of a prompt 3562.84(14)~keV transition.
An intensity upper limit of 1.5\% was derived for the prompt component of the 3562.84(14)~keV transition relative to its stopped component, based on the Compton background intensity.  This upper limit is 4 times lower than the observed prompt feeding from the 842.1(3)~keV transition alone, and 35 times lower than the total intensity of all observed feeders of the 5505.2(2)~keV level.  Based on the large amount of prompt feeding to this level and the lack of a prompt component in its decay, we conclude that the 5505.2(2)~keV level is isomeric and assign it $J^{\pi} = 5^-$.

\begin{figure}
\begin{center}
\includegraphics[width=1.0\columnwidth]{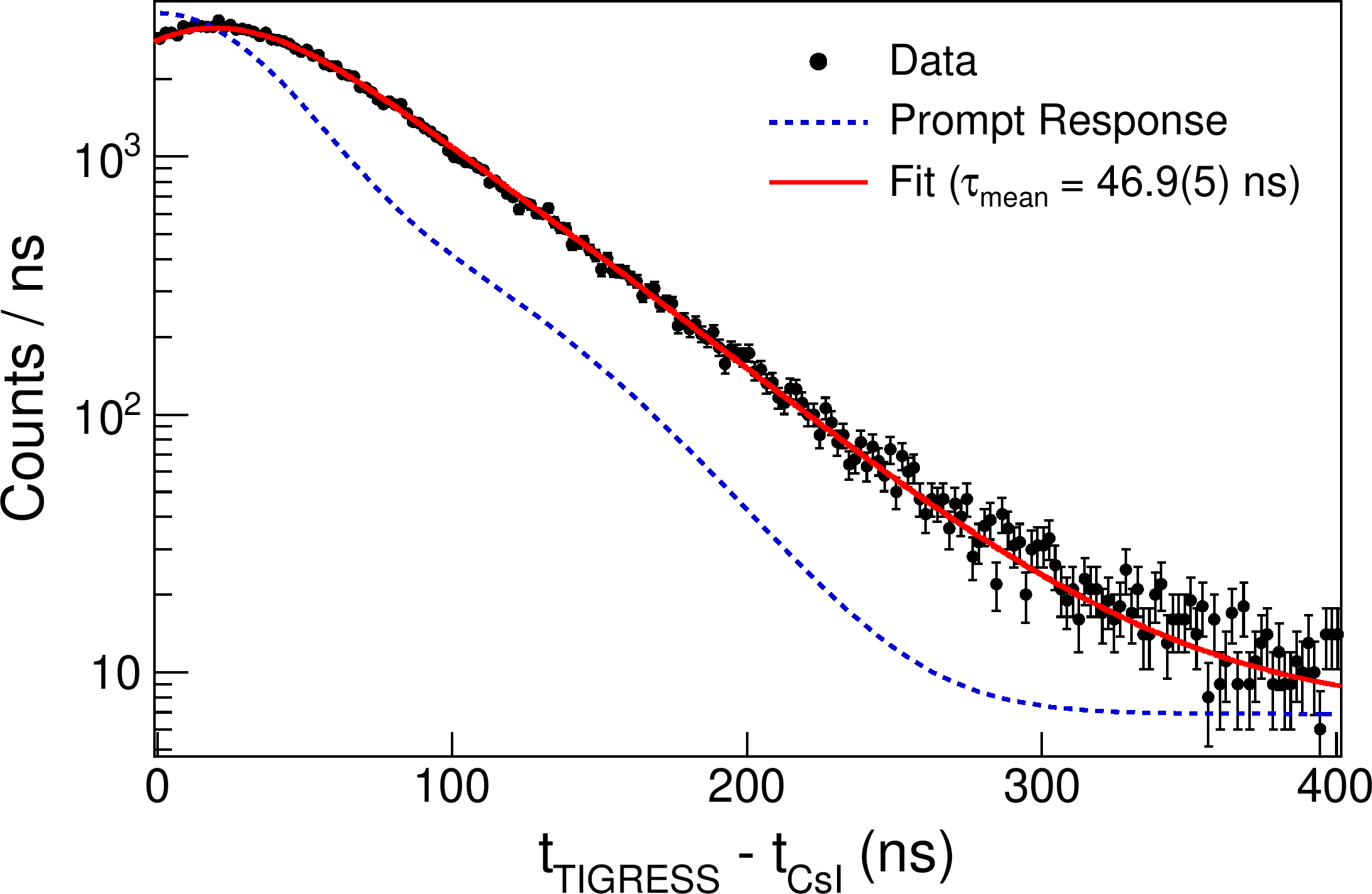}
\end{center}
\caption{Background subtracted TIGRESS-CsI timing distribution for the 3562.84(14)~keV transition in $^{32}$Si.}
\label{fig:tt_3560}
\end{figure}

\par The lifetime of this isomer was determined from the TIGRESS-CsI timing response, shown in Figure \ref{fig:tt_3560}.  The prompt response consisted of background data in the energy range $\pm 150$ keV around the 3562.84(14)~keV transition of interest, and was fit to a double Gaussian function to account for asymmetry of the timing peak.  The full width at half maximum of the prompt response was 78.8(4) ns.  As this width was larger than the expected lifetime of the isomer, the isomer timing response was fit to the prompt response convoluted with an exponential decay function as described in Ref. \cite{regis_generalized_2013}.  The best fit lifetime was $\tau_{mean} = 46.9(5)$ ns, consistent with the value of 47.8(7) ns reported in Ref.~\cite{32si_asai}.  The isomer lifetime was also measured using $\gamma-\gamma$ timing from the 842.1(3) and 3562.84(14)~keV transitions, yielding a result of 44(4) ns, with the larger error in this measurement due to the lower statistics available following $\gamma$-gating.

\par Our identification of the isomeric state contradicts the $4^+_1$ assignment of Ref. \cite{32si_fornal}, with no other $4^+_1$ candidates previously identified.  The yrast $4^+$ state plays an important role in the decay of the isomer as it is one of the few states predicted to have both similar excitation energy and spin.  We have identified a strong candidate for this level at 5881.4(13)~keV, depopulated by an intense 3938.9(13)~keV transition to the $2^+_1$ level.  The DCO ratio and positive $\Delta_{asym}$ value obtained for this transition strongly favour an \emph{E2} assignment.  Based on the high intensity of this transition relative to other lines and the fusion-evaporation reaction used in this experiment, the parent 5881.4(13)~keV level is likely yrast or near-yrast, implying $J > 2$.  We therefore assign $J^{\pi} = 4^+$ to this level.  This is almost certainly the yrast $4^+$ state, and its higher excitation energy compared to the $5^-$ state confirms the existence of a yrast trap in $^{32}$Si.

\begin{figure}
\begin{center}
\includegraphics[width=1.0\columnwidth]{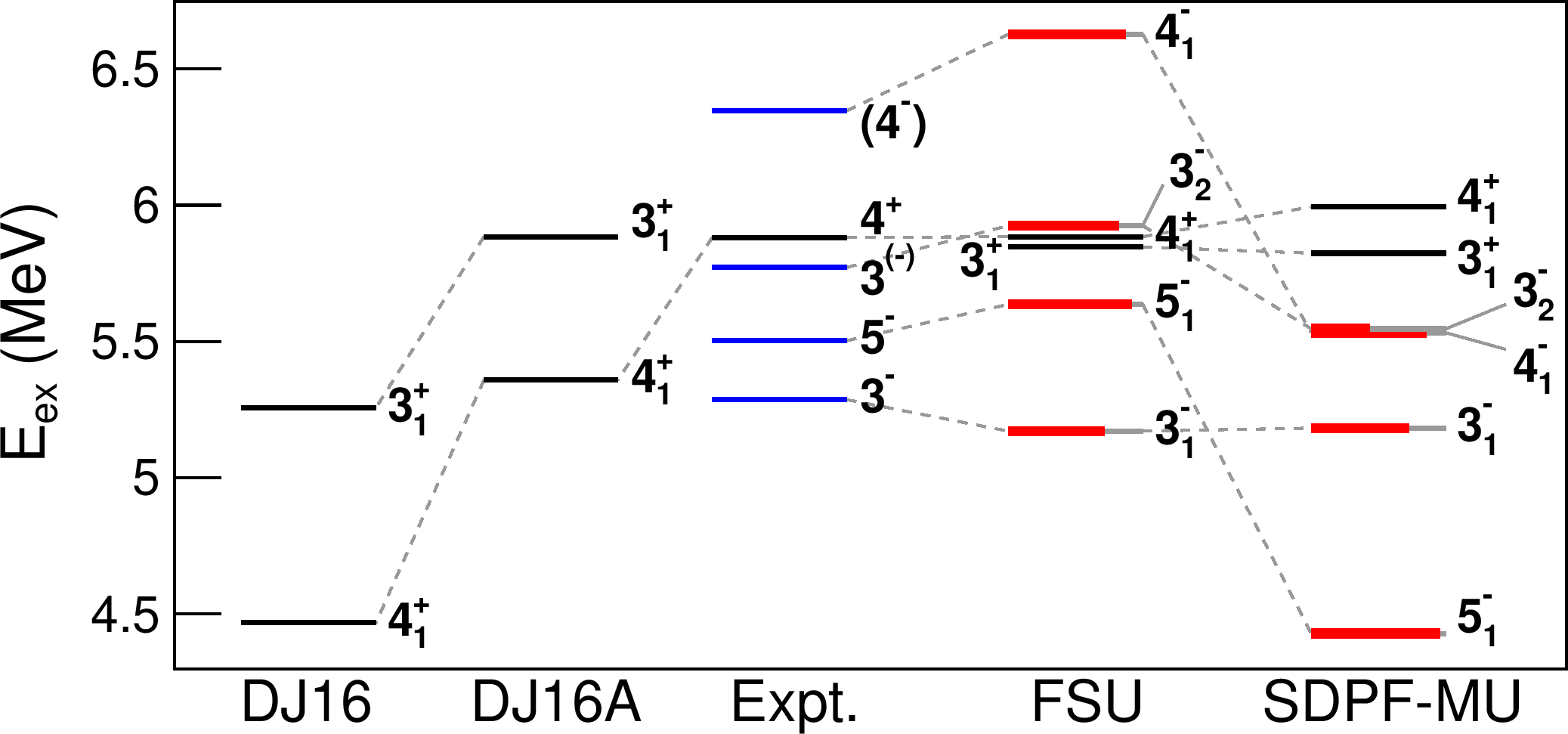}
\end{center}
\caption{Comparison of observed level energies to shell model calculations.  For negative parity states, red bars indicate the calculated $\nu 0f_{7/2}$ orbital population.}
\label{fig:fsu_levels}
\end{figure}

\par This new experimental data significantly clarifies the level structure of $^{32}$Si at intermediate excitation energies sensitive to the effects of cross-shell excitation.  To interpret these findings, shell model calculations were performed using the FSU interaction \cite{lubna2020evolution} in the $psdpf$ valence space, and the SDPF-MU interaction \cite{sdpfmu} in the $sdpf$ valence space.  Negative parity $1p1h$ (1 particle - 1 hole) states were restricted to single neutron excitation in the SDPF-MU calculations, while positive parity states were restricted to the $sd$ shell for both models.
A comparison of level energies is shown in Figure \ref{fig:fsu_levels}.  In general, the energies of positive parity states are similar between both models and consistent with experiment.  For negative parity states, the SDPF-MU calculations appear to underpredict the energies of states with high neutron $0f_{7/2}$ occupancy, while the FSU calculations show better agreement.  This agreement is unsurprising since the relevant two body matrix elements of the FSU interaction are fitted to data in this mass region, with particular focus on negative parity states containing excitation to the $0f_{7/2}$ orbital \cite{lubna2020evolution}.   %This is particularly true for the $5^-_1$ and $4^-_1$ states, where the predicted neutron $0f_{7/2}$ occupancy is highest.  
Also shown in Figure \ref{fig:fsu_levels} are {\it ab-initio} no-core shell model calculations using the DJ16 microscopic effective $sd$ shell interaction \cite{dj16} and the DJ16A interaction which adds a phenomenological monopole modification to DJ16 \cite{dj16a}, both constructed using the Okubo-Lee-Suzuki similarity transformation method \cite{ols}.  The experimental data is better reproduced using the DJ16A interaction, with calculated $2^+_1$ and $4^+_1$ energies that are slightly lower than the experimental values, consistent with the results obtained for lighter $sd$ shell nuclei \cite{choudhary_si_p}.

\par The general agreement of the observed $4^+_1$ energy with many model calculations across various valence spaces suggests that cross-shell excited configurations do not play a significant role in this state.
In the FSU and SDPF-MU calculations, the occupancy of the $\pi s_{1/2}$ and $\pi d_{3/2}$ orbitals is significantly higher for the $4^+_1$ level compared to the lower-lying states, suggesting that the $4^+_1$ energy is sensitive to the $Z=14$ subshell closure.  All models predict $B(E2;~4^+_1 \rightarrow 2^+_1)$ values consistent with our DSAM measurement, see Table \ref{tab:32si_be2_table}.

\par The isomer lifetime measurement yielded a $B(E3;~5^-_1 \rightarrow 2^+_1$) value of $0.0841(10)$ W.u., which is significantly hindered compared to any known \emph{E3} transition in this mass region.  The closest analogue is a similarly hindered $5^-_1 \rightarrow 2^+_1$ \emph{E3} transition in $^{68}$Ni, which has been attributed to its $5^-_1$ state being dominated by the $(\nu p_{1/2})^{-1} \otimes (\nu g_{9/2})^1$ configuration from $N=40$ cross-shell excitation \cite{68Ni_subshell}.
In $^{32}$Si, a similar hindrance of the \emph{E3} strength seems to arise with the dominant $(\nu d_{3/2})^{-1} \otimes (\nu f_{7/2})^{1}$ configuration from $N=20$ cross-shell excitation.  The $5^-_1$ states calculated with FSU and SDPF-MU contain $\nu 0f_{7/2}$ occupancy factors of 0.92 and 0.96, respectively.  The experimental $B(E3;~5^-_1 \rightarrow 2^+_1)$ value is approximately halfway between the values calculated using these models.  The differing configurations of the $5^-_1$ and $2^+_1$ states sufficiently hinder this $E3$ transition such that the $5^-_1$ state is isomeric despite its unusually large $\gamma$-decay energy.  A similar hindrance is evident for the unobserved $5^-_1 \rightarrow 3^-_1$ branch as well as the unobserved $(4^-_1) \rightarrow 4^+_1$ and $(4^-_1) \rightarrow 3^-_1$ transitions, with the FSU and SDPF-MU calculations predicting significantly lower $\nu 0f_{7/2}$ occupancy for the $3^-_1$ state compared to the $4^-_1$ and $5^-_1$ states, as shown in Figure \ref{fig:fsu_levels}.

\begin{table}
\caption{Transition strength values measured in the present work, compared to model calculations.  Effective charges $e_p = 1.36e$ and $e_n = 0.45e$ were used with all models.}
\centering
\begin{ruledtabular}
\begin{tabular}{lllllll}
\rule{0pt}{2.5ex}\multirow{2}{*}{$J^{\pi}_i \rightarrow J^{\pi}_f$} & \multirow{2}{*}{$\lambda L$}  & \multicolumn{5}{c}{$B(\lambda L)$ (W.u.)} \\ \cline{3-7}
\rule{0pt}{2.5ex} & & Expt. & FSU & SDPF-MU & DJ16 & DJ16A  \\ \hline
\rule{0pt}{2.5ex}$2^+_1 \rightarrow 0^+_1$ & \emph{E2} & 6.3$^{+1.1}_{-0.8}$ & 7.3 & 6.3 & 12.4 & 10.1 \\
\rule{0pt}{2.5ex}$4^+_1 \rightarrow 2^+_1$ & \emph{E2} & 8$^{+7}_{-3}$ & 11.2 & 8.2 & 15.2 & 11.4 \\
\rule{0pt}{2.5ex}$5^-_1 \rightarrow 3^-_1$ & \emph{E2} & $< 0.053^{*}$ & 0.077 & 0.6 & - & - \\
\rule{0pt}{2.5ex}$5^-_1 \rightarrow 2^+_1$ & \emph{E3} & 0.0841(10) & 0.150 & 0.012 & - & - \\
\end{tabular}
\end{ruledtabular}
\label{tab:32si_be2_table}
{\footnotesize \rule{0pt}{2.5ex}$^*$From intensity limit of unobserved transition.}
\end{table}

\par
\begin{figure}
\begin{center}
\includegraphics[width=1.0\columnwidth]{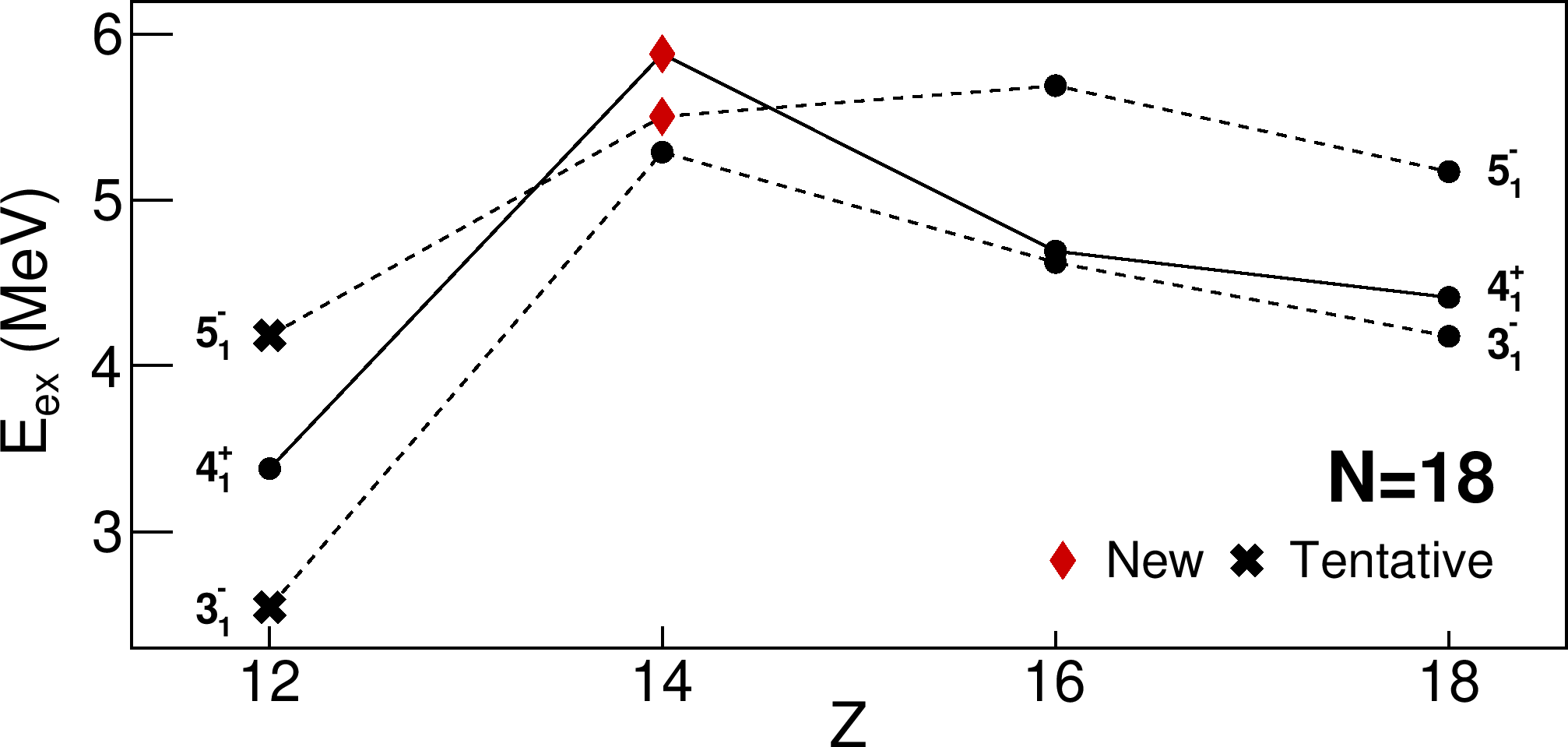}
\end{center}
\caption{Systematics of yrast states for $N=18$ isotones in the vicinity of $^{32}_{14}$Si.  Data from Refs.~\cite{nds30,nds32,nds34,nds36} and this work.}
\label{fig:systematics}
\end{figure}

 Figure \ref{fig:systematics} shows a comparison of the present data to studies of other $N=18$ isotones, indicating that the re-ordering of yrast states is specific to $^{32}_{14}$Si.  The $4^+_1$ level energy is enhanced at the $Z=14$ subshell closure, which has been shown to widen with increasing neutron number \cite{bespalova_evolution_2018}.  For the negative parity states, both the FSU and SDPF-MU calculations show that the narrowing of the $N=20$ shell gap becomes significant for $Z < 14$, and the FSU calculations show that mixed proton-neutron excitation to $0f_{7/2}$ becomes significant for $Z > 14$ (since proton excitation to $0f_{7/2}$ becomes more energetically favorable with increased ground state occupation of the upper $sd$ orbitals).  Both of these effects lower the energy of the $3^-_1$ and $5^-_1$ states, with the latter effect being stronger for the $3^-_1$ states.  At $Z=14$, the proton $0f_{7/2}$ contribution is mostly absent, with the FSU calculations showing an occupancy of 0.074 for the $3^-_1$ state in $^{32}_{14}$Si, compared to 0.325 in $^{34}_{16}$S.  This reduces the energy gap between the $3^-_1$ and $5^-_1$ states in $^{32}$Si which suppresses the $5^-_1 \rightarrow 3^-_1$ decay branch.
The spin trapped $5^-_1$ isomer therefore arises due to the varied effects of proton and neutron cross-shell excitation on different states in the yrast sequence.

\par In summary, the properties of the $^{32}$Si isomeric state were determined, with the $4^+_1$ state identified at higher energy.  This shows that the isomer is located in a yrast trap, which forms due to the relevant yrast states containing different configurations sensitive to proton and/or neutron cross-shell excitation.  This is the only known case of a spin isomer in an even-even $sd$ shell nucleus, and a rare example of an isomer with a very high $\gamma$-decay energy.  High-spin isomers are best populated using fusion-evaporation reactions, however many neutron rich nuclides have yet to be studied in this way due to the challenge imposed by low reaction cross sections.  Development of higher intensity radioactive beams will help to address this challenge.

\begin{acknowledgments}
\par The authors appreciate the support of the ISAC Operations Group at TRIUMF and the Simon Fraser University Electronics and Machine Shops.  This work was supported by the Natural Sciences and Engineering Research Council of Canada, the Canadian Foundation for Innovation and the British Columbia Knowledge Development Fund.  TRIUMF receives federal funding through a contribution agreement with the National Research Council of Canada.   This work was also supported by the U.S. Department of Energy, Office of Science, Office of Nuclear Physics under Award No.~DE-SC0020451 (FRIB), and by the U.S. Department of Energy Grant No.~DE-FG02-93ER40789.  FSU shell model calculations were performed using the computational facility of Florida State University, supported by Grant No.~DE-SC0009883 (FSU).  P.C.S.~acknowledges financial support from SERB (India), CRG/2022/005167.
\end{acknowledgments}

%\bibliography{bibfile}% Produces the bibliography via BibTeX.

%merlin.mbs apsrev4-1.bst 2010-07-25 4.21a (PWD, AO, DPC) hacked
%Control: key (0)
%Control: author (8) initials jnrlst
%Control: editor formatted (1) identically to author
%Control: production of article title (-1) disabled
%Control: page (0) single
%Control: year (1) truncated
%Control: production of eprint (0) enabled
%

\end{document}